\documentclass[journal]{IEEEtran}

\usepackage[bookmarks=false]{hyperref}
\usepackage{url}
\usepackage{array}
\newcolumntype{C}[1]{>{\centering\arraybackslash}m{#1}}

\ifCLASSINFOpdf
   \usepackage[pdftex]{graphicx}
\fi

\begin{document}
\title{$\mu$TCA DAQ system and parallel reading in CANDLES experiment}

\author{B. T. Khai,~\IEEEmembership{Member,~IEEE,} 
		S. Ajimura,~\IEEEmembership{Member,~IEEE,} 
		K. Kanagawa, 
		T. Maeda, \\
		M. Nomachi,~\IEEEmembership{Senior Member,~IEEE,} 
		Y. Sugaya, 
		K. Suzuki, 
		and M. Tsuzuki 
\thanks{The article is submitted on June 24, 2018.
		This work was presented at The 21st RealTime Conference in Williamsburg, Virginia, 2018. 
		This work is supported by CANDLES collaborators. }
\thanks{B. T. Khai, K. Kanagawa, T. Maeda, and M. Tsuzuki are with Graduate School of Science, Osaka University, Toyonaka, Osaka 560-0043, Japan.}
\thanks{S. Ajimura, M. Nomachi, and Y. Sugaya are with Research Center for Nuclear Physics, Osaka University, Ibaraki, Osaka 567-0047, Japan.}
\thanks{K. Suzuki is with The Wakasa Wan Energy Research Center, 64-52-1 Nagatani, Tsuruga, Fukui 914-0135, Japan.}
}

\markboth{
21st IEEE Real Time Conference, DAQ section, Oral Presentation, Contribution ID: 590
}%
{
}

\maketitle

\begin{abstract}
A new $\mu$TCA DAQ system was introduced in CANDLES experiment with SpaceWire-to-GigabitEthernet (SpaceWire-GigabitEthernet) network for data readout and Flash Analog-to-Digital Converters (FADCs). 
With SpaceWire-GigabitEthernet, we can construct a flexible DAQ network with multi-path access to FADCs by using off-the-shelf computers. 
FADCs are equipped 8 event buffers, which act as de-randomizer to detect sequential decays from the background. 
SpaceWire-GigabitEthernet has high latency (about 100 $\mu$sec) due to long turnaround time, while GigabitEthernet has high throughput. 
To reduce dead-time, we developed the DAQ system with 4 ``crate-parallel'' (modules in crates are read in parallel) reading threads. 
As a result, the readout time is reduced by 4 times: 40 msec down to 10 msec. 
With improved performance, it is expected to achieve higher background suppression for CANDLES experiment. 
Moreover, for energy calibration, ``event-parallel'' reading process (events are read in parallel) is also introduced to reduce measurement time. 
With 2 ``event-parallel'' reading processes, the data rate is increased 2 times.
\end{abstract}


\ifCLASSOPTIONpeerreview
\centering \bfseries EDICS Category: 3-BBND 
\fi
%
\IEEEpeerreviewmaketitle

\section{CANDLES experiment}
\IEEEPARstart{N}{eutrino-less} Double Beta Decay (0$\nu\beta\beta$) phenomenon is an important tool to study absolute neutrino mass and nature of neutrino (Majorana or Dirac particle). 
On top of that, this phenomenon signals the violation of lepton number conservation. 
Thus, by studying this phenomenon, we can explore the new physics beyond Standard Model. 
CANDLES (Calcium fluoride for studies of Neutrino and Dark matters by Low Energy Spectrometer) is searching for 0$\nu\beta\beta$ from $^{48}$Ca. 
It is constructed in Kamioka Underground Observatory (2700 m.w.e depth). 
This is big challenge due to extremely rare decay rate (T$_{1/2}^{0\nu\beta\beta} > 5.8\times10^{22}$ years \cite{ELEGANT}). 
Thus, to obtain 0$\nu\beta\beta$, it requires a large amount of source and low background environment.
\par
\begin{figure}
\centering
\includegraphics[width=55mm]{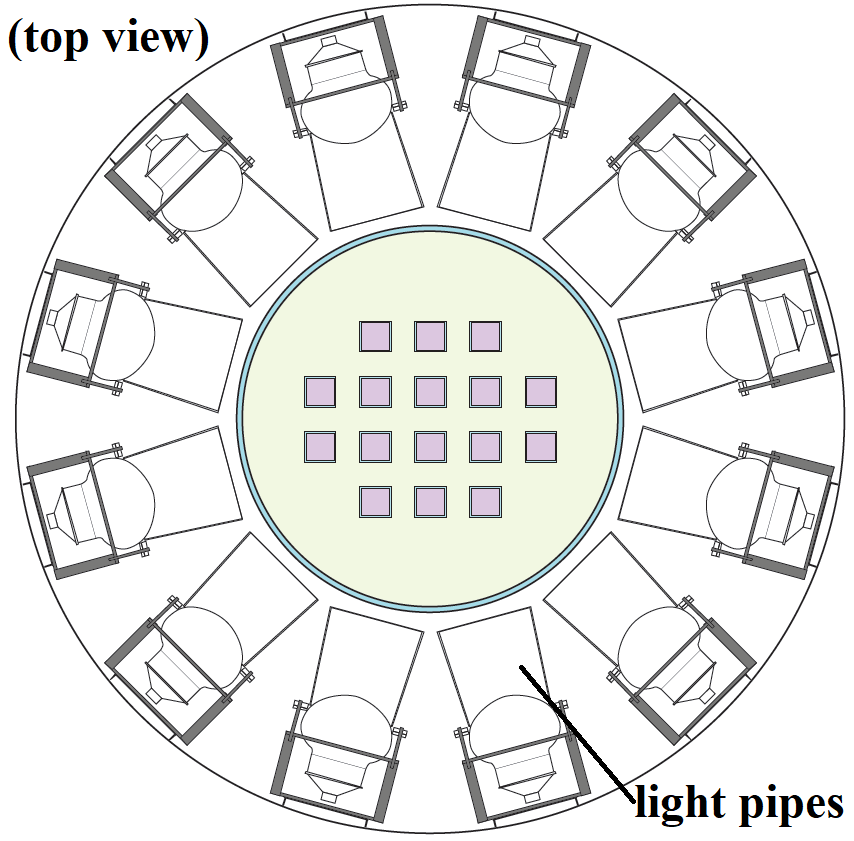}
\includegraphics[width=55mm]{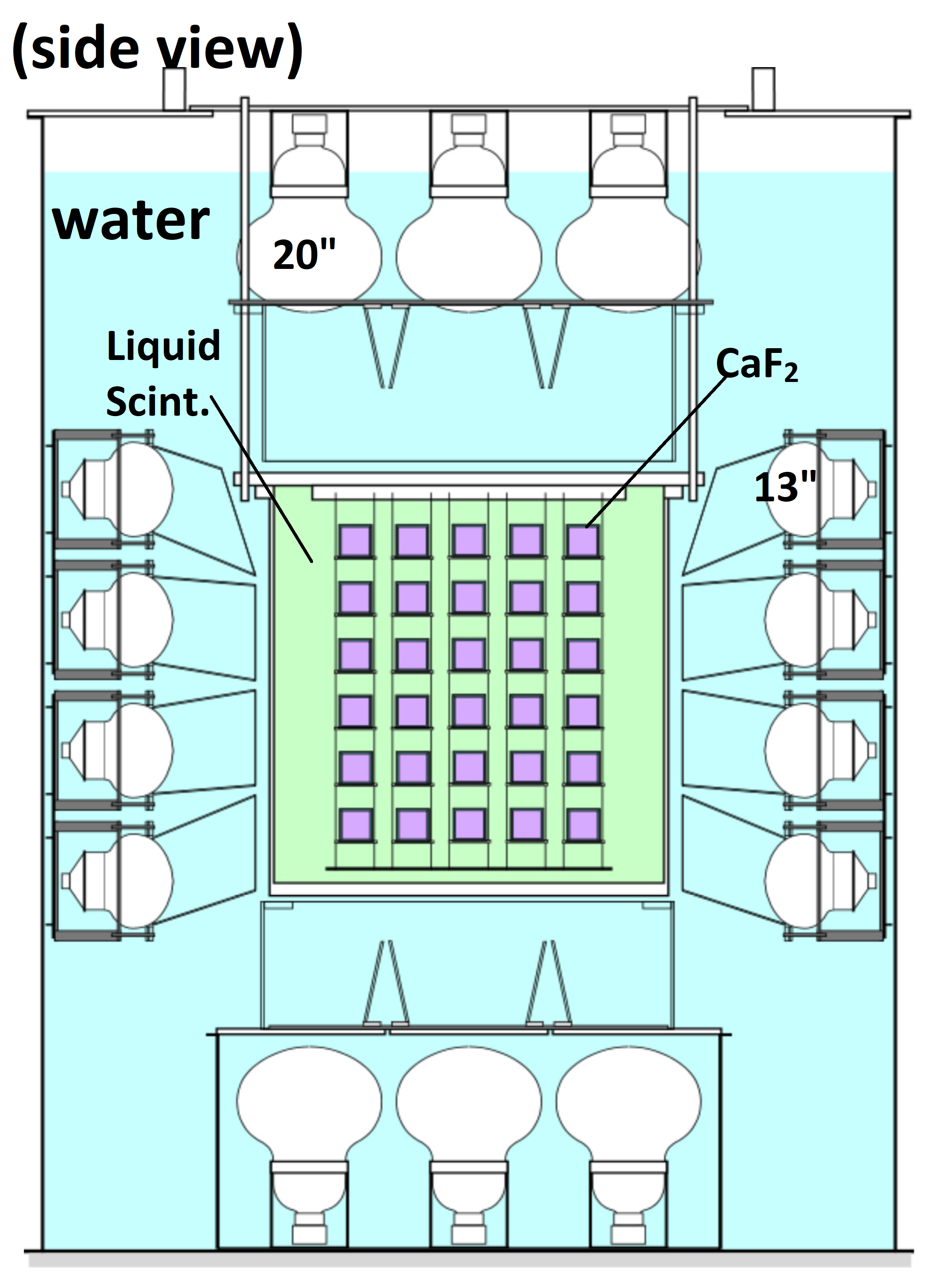}
\caption{Experimental setup of CANDLES III setup with top view and side view. CaF$_{2}$ crystals are submerged in a vessel of Liquid scintillator. Liquid scintillator vessel and PMTs are mounted inside a cylindrical water tank made from stainless steel. Light pipes are placed between PMTs and LS vessel to increase photo-coverage. }
\label{figure_CANDLES_setup}
\end{figure}

Schematic view of CANDLES experimental setup is shown in Figure \ref{figure_CANDLES_setup}. 
CANDLES uses 96 cubic crystals of CaF$_{2}$ with dimensions of 10 cm. 
These crystals are submerged inside 2000 liters of liquid of scintillator (LS). 
Decay constant of LS and CaF$_{2}$ waveforms are $\textit{O}$(10 nsec) and $\textit{O}$(1 $\mu$sec), respectively. 
LS is used as an active veto in CANDLES due to the difference of pulse shapes. 
Thus, we use high-performance Flash Analog-to-Digital Converters (FADCs) and Pulse Shape Discrimination to remove the background. 
Scintillator photons are detected by 62 photomultiplier tubes (PMTs) surrounding. 
They consist of forty-eight 13-inch PMTs (R8055, Hamamatsu) on the side and fourteen 20-inch PMTs (R7250, Hamamatsu) on top and bottom. 
Light-concentration system including light pipes is placed between LS vessel and PMTs to increase photo-coverage \cite{CANDLES_Umehara_EPJ_2014}. 
In order to increase the light output of CaF$_{2}$ and collection efficiency of PMTs, we install the magnetic cancellation system and keep detector running at 2 $^{o}$C \cite{CANDLES_IIDA_Taup}. 
Everything is mounted in a stainless-steel cylindrical water tank with 4 m of height and 3 m of diameter. 

Q-value of $^{48}$Ca (4.3 MeV) is highest among all double beta decay isotopes and far from most of natural background. 
Although most of background is removed by active shielding, there are still background events around the energy of interest. 
These background events contain: 
\begin{itemize}
\item[(a)] Two-neutrino double beta decay (2$\nu\beta\beta$)
\item[(b)] External background from (n,$\gamma$) reactions
\item[(c)] Impurities background:
    \begin{itemize}
    \item[(c-1)] BiPo sequential decays
    \item[(c-2)] $\beta$-decay events $^{208}$Tl
    \end{itemize}
\end{itemize}
For 2$\nu\beta\beta$ background (a), we need to improve energy resolution to discriminate 0$\nu\beta\beta$ from 2$\nu\beta\beta$. 
The (n,$\gamma$) reactions inside and outside the water tank induce background around our energy of interest (b). 
We need to install passive shielding including lead (to reduce $\gamma$-rays) and boron (to reduce neutron). 
The BiPo sequential decays (c-1) are of $^{212}$Bi-$^{212}$Po and $^{214}$Bi-$^{214}$Po. 
The sequential decays from Bi and Po form a pile-up event acting as a background event in CANDLES. 
Since they are pile-up events, we can reject these events by Pulse Shape Discrimination. 
In case $\beta$-decays from of $^{208}$Tl, they have high Q-value (Q$_{\beta+\gamma}$ $\approx$ 5 MeV) and occur after $\alpha$-decay of $^{212}$Bi. 
To remove this $\beta$-decay, we tag the preceding $\alpha$-decay. 
For high tagging efficiency, dead-time of the DAQ system must be minimized at CANDLES trigger rate (20 cps \cite{IEEE-Suzuki}). 
This is the demanding work for the DAQ system.

\section{DAQ SYSTEM OF CANDLES}
The $\mu$TCA (Micro Telecommunications Computing Architecture) system had been developed since 2010 in cooperation with JAXA (Japan Aerospace Exploration Agency) \cite{JAXA} and Shimafuji Electric Co. \cite{Shimafuji}. 
In 2016, we introduced $\mu$TCA DAQ system in CANDLES with SpaceWire-to-GigabitEthernet (SpaceWire-GigabitEthernet) and 8 event buffers. 
Multiple event buffers and SpaceWire-GigabitEthernet network are explained in section \ref{AMC-FADCs} and section \ref{SpaceWire}, respectively. 
Since we started the development before DESY released MTCA.4, it does not follow MTCA.4 standard \cite{MTCA.4}. 
Details of the system are indicated in Table \ref{table_uTCA}. 
\subsection {$\mu$TCA DAQ and development of MicroTCA Carrier Hub}
\begin{figure}
\centering
\includegraphics[width=85mm]{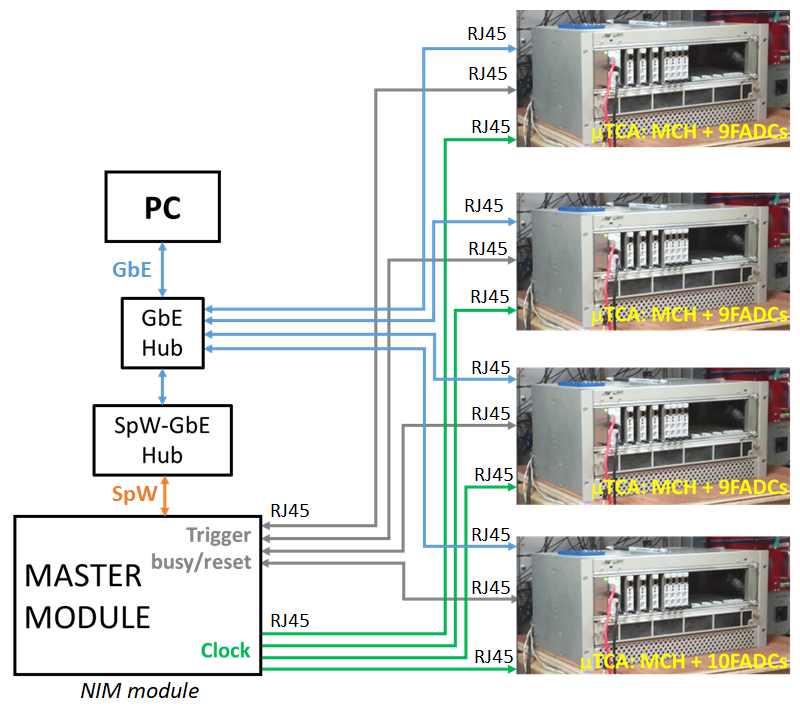}
\caption{$\mu$TCA system including MCH and AMC-FADCs used in CANDLES. Master Module is introduced to distribute clock and trigger signals. PC can access via SpaceWire-to-GigabitEthernet interface to SpaceWire network, where SpaceWire modules (AMC-FADCs, Trigger FPGAs, Master Module) are connected.}
\label{figure_DAQ_diagram}
\end{figure}

Schematic diagram of our $\mu$TCA DAQ is shown in Figure \ref{figure_DAQ_diagram}. 
There are 37 Advanced Mezzanine Cards FADC (AMC-FADC) modules. Each AMC-FADC has 2 channels. 
These AMC-FADCs are housed in four $\mu$TCA crates. 
In each crate, there are one MicroTCA Carrier Hub (MCH) and 9 (or 10) AMC-FADCs. 
The MCH can access all AMC-FADCs in the same crate via the backplane. 
Master Module, which is developed in NIM (Nuclear Instrument Module) standard, is introduced to distribute global clock and global trigger signals. 

\begin{table}[!t]
\renewcommand{\arraystretch}{1.3}
\caption{Details of $\mu$TCA system used in CANDLES}
\label{table_uTCA}
\centering
\begin{tabular}{C{1.3cm}|C{1.6cm}|C{2.3cm}|C{2cm}}
\hline
\hline
\bfseries Module & \bfseries Manufacturer & \bfseries FPGA & \bfseries FPGA \\
& & \bfseries logic & \bfseries development \\
\hline
$\mu$TCA & Uber Ltd. & & \\
\hline
MCH & Shimafuji & GbE-SpW interface & by Shimafuji \\
 & & Clock distribution & \\
\cline{3-4}
 & & SpaceWire Router & Open IP\\
 & & & by Shimafuji \\
\cline{3-4}
 & & Trigger Controller & by Osaka\\
 & & for CANDLES & University\\
\hline
AMC-FADC & Shimafuji & FADC control & by RCNP, \\
 & & & Osaka University\\
\cline{3-4}
 & & SpaceWire & Open IP \\
 & & & by Shimafuji \\
\hline
\hline
\end{tabular}
\end{table}

\begin{figure}
\centering
\includegraphics[width=90mm]{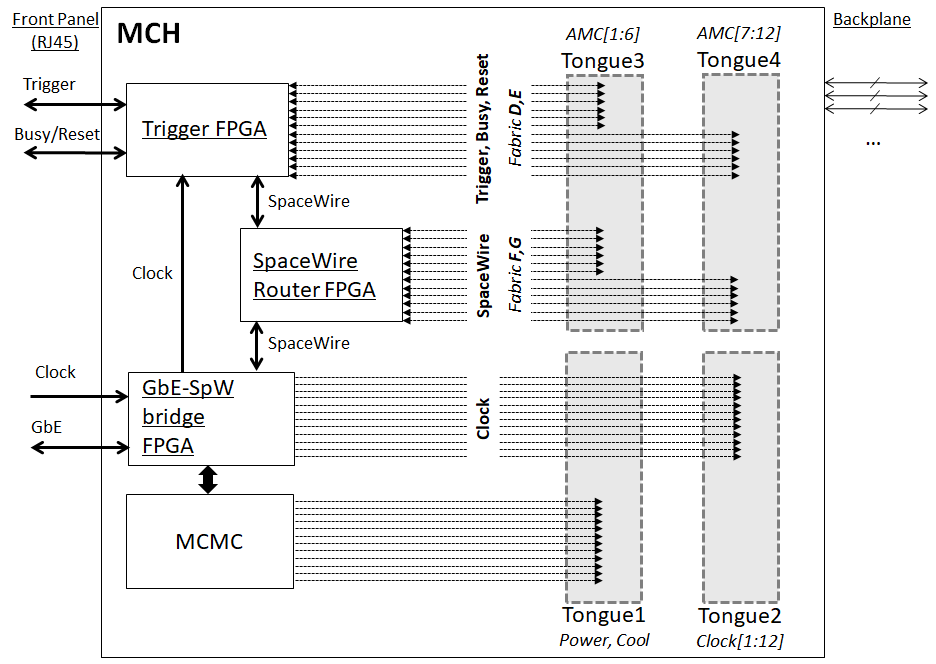}
\caption{MCH is developed for CANDLES experiment. It contains 3 FPGAs: GbE-SpW bridge (GbE-to-SpW interface and clock distribution), SpaceWire Router (access Trigger FPGA and all AMC-FADCs via backplane) and Trigger FPGA (collect/distribute local trigger, busy, reset signals from/to AMC-FADCs).}
\label{figure_MCH}
\end{figure}

Figure \ref{figure_MCH} shows the diagram of an MCH. 
The MCH has three Spartan6 XC6SLX100 FPGAs (Field Programmable Gate Arrays) \cite{Xilinx}: GigabitEthernet-SpaceWire bridge FPGA, Trigger FPGA, and SpaceWire Router FPGA. 
These 3 FPGAs of MCH are used for data readout via SpaceWire-GigabitEthernet network, clock distribution, collection of local trigger signals and broadcasting global trigger signal:
\begin{itemize}
    \item Data Readout: All SpaceWire modules are connected to SpaceWire Routers in MCHs. In data readout, PC accesses from GigabitEthernet network to SpaceWire network, where all SpaceWire modules are connected, via GigabitEthernet-SpaceWire interface. PC can set parameters and read data. 
    \item Clock Distribution: A 25 MHz common clock is distributed from Master Module to GigabitEthernet-SpaceWire FPGAs in 4 MCHs. The GigabitEthernet-SpaceWire FPGA forwards the clocks to Trigger FPGA and all AMC-FADCs. When arriving at FPGA (Cyclone 4 EP4CE30 \cite{Cyclone}) in AMC-FADC, the clock is manipulated to 125 MHz by PLL (Phase-locked loop) for readout buffer and 500 MHz by Clock Synthesizer (MAX3674 - Microsemi \cite{Microsemi}) for 500 MHz FADC (ADC08DL502 - Texas Instrument \cite{TI}). 
    \item Trigger Control: Every AMC-FADC has local trigger decision to select CaF$_{2}$ events using Dual Gate Trigger \cite{IEEE-Maeda}. When there is CaF$_{2}$ event, AMC-FADC sends the local trigger to Trigger FPGA in MCH. Trigger FPGA forwards the local trigger to Master Module. After that, Master Module broadcasts the global trigger to all Trigger FPGAs. Trigger FPGA sends the global trigger to all AMC-FADCs via the backplane. When the global trigger comes, the waveform will be recorded in an event buffer of AMC-FADC. 
\end{itemize}

\subsection{Backplane connection}
\begin{figure*}
\centering
\includegraphics[width=180mm]{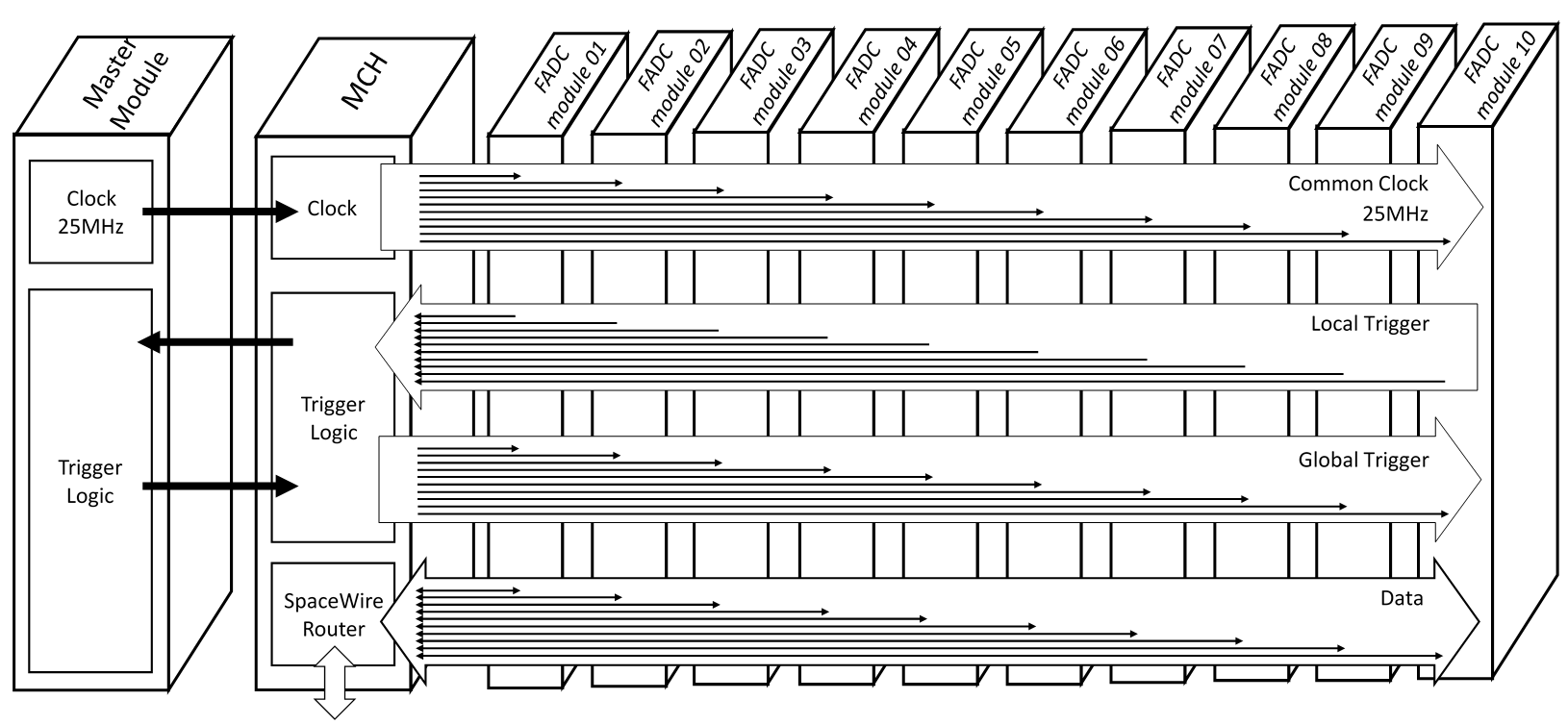}
\caption{Tripple-stars network in one $\mu$TCA crate. One star is used to connect trigger module with AMC-FADC modules. On star is used for common clock distribution. The other one is for connecting SpaceWire router with AMC-FADCs.}
\label{figure_MTCA_system}
\end{figure*}

The connection between MCH and AMC-FADCs are done via backplane of $\mu$TCA crate. 
The $\mu$TCA is a point-to-point star data links. 
Figure \ref{figure_MTCA_system} shows the triple-stars in one $\mu$TCA crate. 
One star connection is used for common clock distribution. 
The common 25 MHz clock is distributed from Master Module. 
One star connection is used to connect trigger module with AMC-FADCs. 
The global trigger is broadcast from Master Module after receiving local trigger residing from AMC-FADCs. 
The last one is used to connect SpaceWire router and AMC-FADCs. 
PC can access to AMC-FADCs via this SpaceWire router and SpaceWire-to-GigabitEthernet interface. 
\par
Table \ref{table_backplane} indicates port mapping between MCH and AMC-FADCs for trigger controlling, SpaceWire router, clock distribution and power/cooling management. 
In the backplane, we assign two ports for SpaceWire, one port for trigger and one port for busy and reset. 
The GigabitEthernet-SpaceWire bridge FPGA has a GigabitEthernet-to-SpaceWire interface. 
This FPGA has access to Tongue 2 to distribute common clock to AMC-FADCs. 
Trigger FPGA has access to Fabric D and E in Tounge 3 and Tongue 4 to send (and receive) trigger, busy and reset signals. 
These signals can approach every AMC-FADC via AMC ports 4 and 5 in Fat pipe: one port is used for trigger signal and the other one is used for busy/reset signal.
GigabitEthernet-to-SpaceWire interface and Trigger FPGA are connected to the SpaceWire Router FPGA. SpaceWire Router FPGA has access to Fabric F and G in Tongue 3 and Tongue 4 to form a SpaceWire network in the backplane. 
The SpaceWire network is connected to every AMC-FADC via AMC ports 6 and 7 in the Fat pipe. 
Since SpaceWire uses 2 ports in backplane, these two AMC ports are used for one SpaceWire connection.

\begin{table}[!t]
\renewcommand{\arraystretch}{1.3}
\caption{Backplane connectors of MCH and AMC-FADC}
\label{table_backplane}
\centering
\begin{tabular}{C{3.5cm}|C{2cm}|C{2cm}}
\hline
\hline
 & \bfseries MCH & \bfseries AMC-FADC \\
\hline
\bfseries Power/Cooling & Tongue 1 & \\
\bfseries Management & & \\
\hline
\bfseries Clock Distribution & Tongue 2 & \\
(common clock) & & \\
\hline
\bfseries Trigger Control & Tongue 3 and 4 & Fat pipes 4 and 5\\
(trigger/busy/reset signals) & (Fabric D and E)& \\
\hline
\bfseries SpaceWire Router & Tongue 3 and 4 & Fat pipes 6 and 7\\
(SpaceWire packets) & (Fabric F and G) & \\
\hline
\hline
\end{tabular}
\end{table}

\subsection{AMC-FADC and multiple event buffers} \label{AMC-FADCs}
The AMC-FADCs with 500 MHz sampling speed, 8-bit resolution, and 2 input channels are developed by Research Center for Nuclear Physics (RCNP, Osaka University) and Shimafuji Electric Co. \cite{Shimafuji}. 
In CANDLES, the trigger decisions are made using waveform of analog sum signals of 62 PMTs. 
Beside 62 channels recording waveform of each PMT, we use 12 channels to record analog sum signals for trigger purposes. 
For waveform recording, signals from all PMTs are fed to AMC-FADCs.
To record long decay-constant signal of CaF$_{2}$, a time window of digitized waveform of all AMC-FADCs are set for about 9 $\mu$sec. 
To reduce data size, we use AMC-FADCs with summation function \cite{IEEE-Nomachi-2006}, \cite{Umehara-IEEE-2011}. 
For first 768 nsec, the waveform is digitized every 2 nsec. 
After 768 nsec, digitized values in 64 nsec are summed up and recorded as 16 bits of data. 
Data in one waveform includes 384 points of 8 bits and 128 points of 16 bits. 
Waveform data of one AMC-FADC channel is 640 Bytes, and the total data size of one event in CANDLES is nearly 50 kBytes. 
\par
\begin{figure}
\centering
\includegraphics[width=80mm]{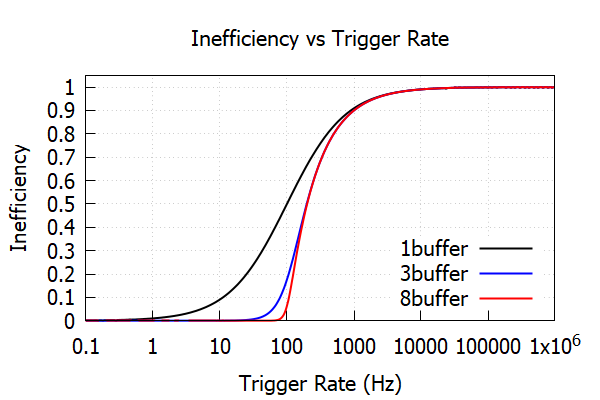}
\caption{The inefficiency of DAQ with 1, 3 and 8 buffers are plotted in black, blue and red color, respectively. In this calculation, the readout time is set at 10 msec. Inefficiency is reduced by using more event buffers.}
\label{figure_InEff_Buffers}
\end{figure}

Random events from background cause inefficiency in the DAQ. 
To reduce dead-time in CANDLES, each AMC-FADC channels is equipped with a ring buffer. 
It contains 8 event buffers which act as derandomizer. 
This ring buffer is installed at the front-end electronics of AMC-FADCs. 
The event buffer works as M/D/1/N queue, which shows random event arrival at N event buffers and deterministic read-time by one process. 
Stationary distribution for the number of events is calculated using recurrence formula. 
Inefficiency occurs when all event buffers are occupied. 
According to queue theory \cite{Queue-Theory}, one can calculate the inefficiency as a function of (readout time$\times$trigger-rate). 
Figure \ref{figure_InEff_Buffers} shows the inefficiency of DAQ system as a function of trigger rate. 
In this calculation, the readout time is set at 10 msec. 
From the figure, we can obtain less inefficiency by using more event buffers.

\section {SpaceWire network in CANDLES DAQ system} \label{SpaceWire}
\subsection {SpaceWire and Remote Memory Access Protocol}
\begin{figure}
\centering
\includegraphics[width=90mm]{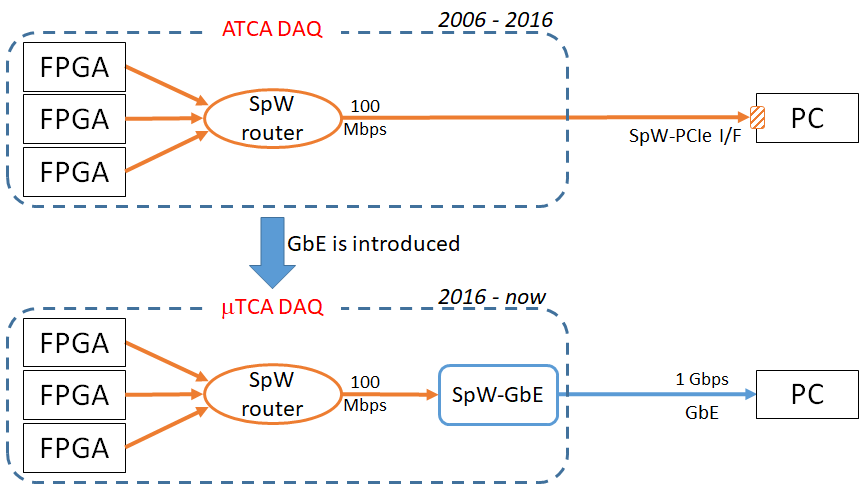}
\caption{SpaceWire network in previous ATCA DAQ system (top) and new $\mu$TCA DAQ system (bottom). In ATCA system, A SpaceWire-to-PCIe interface was used to access to SpaceWire network from PC. In $\mu$TCA system, we introduce SpaceWire-to-GigabitEthernet as an easy interface to PC.}
\label{figure_SpaceWire}
\end{figure}

SpaceWire is developed based on DS-links \cite{DS-links}. 
It connects sub-systems on board spacecraft. 
We use ECSS-E-ST-50-12C SpaceWire Standard \cite{SpaceWire-Standard} in our system. 
It is bi-directional and full-duplex data link. 
Data bandwidth (adjustable from 10 Mbps to 200 Mbps) is set at 100 Mbps in CANDLES. 
The physical layer of SpaceWire is LVDS (Low Voltage Differential Signaling). 
Therefore, it is possible to use $\mu$TCA backplane to connect SpaceWire nodes. 
SpaceWire can be implemented in a small number of logic elements.
All FPGAs have SpaceWire interfaces. 
Registers are accessible from PC via SpaceWire network. 
Figure \ref{figure_SpaceWire} is the comparison of SpaceWire network of previous ATCA DAQ system and new $\mu$TCA DAQ system. 
In previous DAQ system \cite{IEEE-Suzuki} \cite{IEEE-Maeda}, a SpaceWire-to-PCIe (PCI express) interface was used. 
This setup has short latency, but it is not convenient due to two reasons. 
First, it requires special device driver which is not easy to maintain. 
Second, PCIe slot is required, hence, the selection of PCs is limited. 
In the new $\mu$TCA system (this work), we have introduced SpaceWire-GigabitEthernet converter as an interface to PC. 
With Gigabit Ethernet, there is no need to use a special interface and we can use any off-the-shelf computer for data readout. 

\par
\begin{figure}
\centering
\includegraphics[width=60mm]{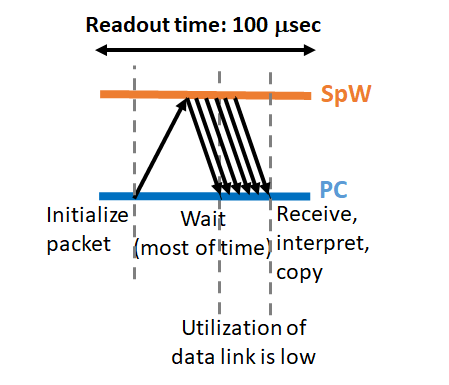}
\caption{At small packet size, readout time (in one transaction) is the SpaceWire-GigabitEthernet network overhead . Most of time is waiting time. Utilization is very low.}
\label{figure_RMAP_transaction}
\end{figure}
\begin{figure}
\centering
\includegraphics[width=80mm]{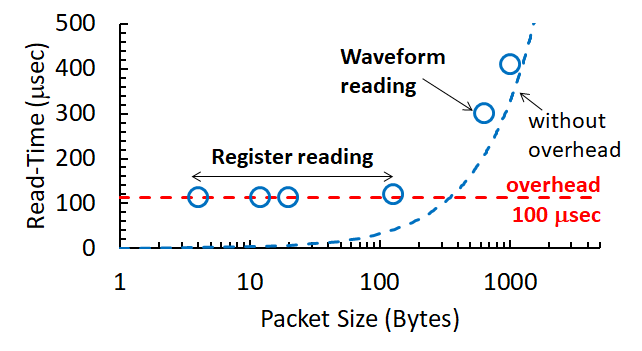}
\caption{Readout time is a function of data size in SpaceWire-GigabitEthernet network. The overhead (about 100 $\mu$sec) is plotted in red dashed line. The blue dashed line is the readout time without the overhead. In CANDLES, we need to read register information (from 4 Bytes to 128 Bytes of packet size) and waveform data (640 Bytes of packet size).}
\label{figure_ReadTime}
\end{figure}

RMAP (Remote Memory Access Protocol) ECSS-E-ST-50-52 standard \cite{RMAP-Standard} is used in our system to access register in front-end modules. 
For communication between PCs and DAQ system, we used SpaceWireRMAPLibrary \cite{SpaceWire-RMAP-lib}. 
It is an open source C++ library for developments of SpaceWire network and data transfer through RMAP. 
RMAP is a transaction of sending request and getting reply (Figure \ref{figure_RMAP_transaction}). 
Front-end FPGA responds quickly.
However, TCP/IP (Transmission Control Protocol/Internet Protocol) latency due to send request and to receive reply cannot be ignored. 
The overhead of SpaceWire-to-GigabitEthernet is about 100 $\mu$sec (Figure \ref{figure_ReadTime}). 
The overhead is mainly the waiting time in RMAP transaction. 
This overhead is dominant in data readout of small-size registers. 
Since we read waveform data and many data from registers, the DAQ speed is limited by large number of accesses to small-size registers. 
Thus, we need to reduce readout time.
Bandwidth utilization is very low. 
Therefore, in order to reduce readout time, we conduct parallel-readout.

\subsection {Parallel readout}
As mentioned in the previous section, there are 74 AMC-FADC channels stored in four $\mu$TCA crates. 
Data of these channels are event fragments. With sequential reading, PC reads all crates one by one to get full data set. 
There are two kinds of parallel readout that we use in CANDLES: crate-parallel readout and event-parallel readout. 
In case of crate-parallel, data in 4 crates are read at the same time. 
These 4 reading processes are shared by 4 threads in the PC. 
We can increase DAQ speed by 4 times. 
Event-parallel readout means events in buffers are read in parallel. 
In this work, we conduct 2 event-parallel readouts to speed up DAQ speed in high trigger rate. 
To do that, 2 readers are implemented for reading two events at the same time: one reads odd event buffer and the other reads even event buffer. 
Thus, with 2 event-parallel, we can increase DAQ speed 2 times faster than non-parallel readout. 
Combing crate-parallel readout, DAQ speed can be increased 8 times. 

\subsection {DAQ-Middleware}
\begin{figure}
\centering
\includegraphics[width=80mm]{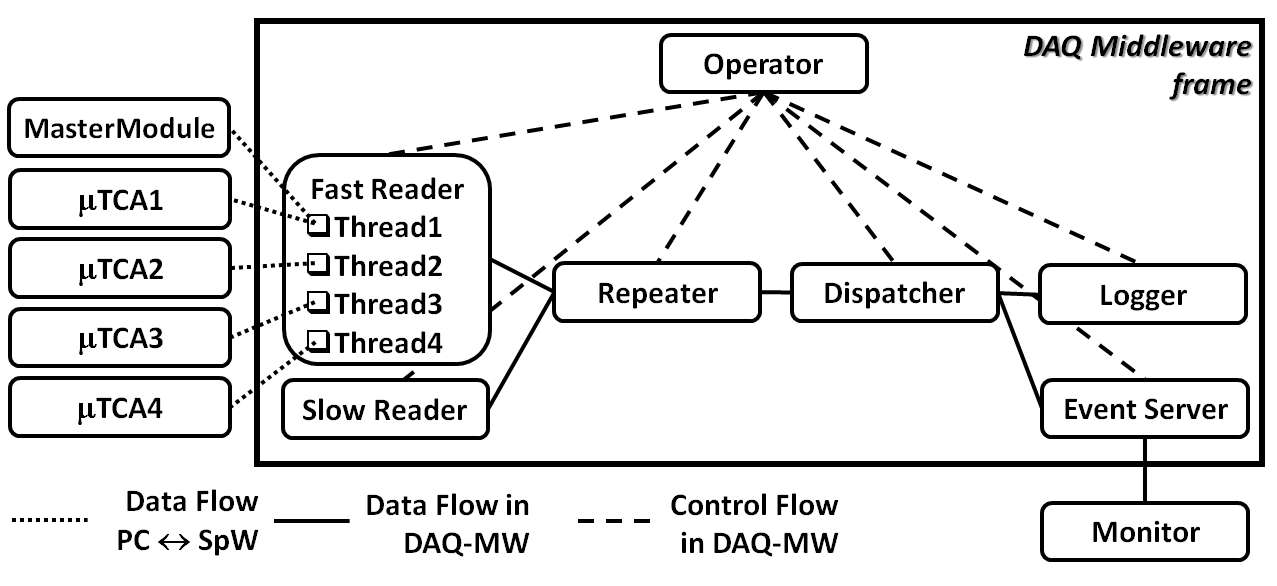}
\caption{Schematic diagram of data flow inside DAQ Middleware. For crate-parallel readout, four threads are implemented inside Fast Reader to read data in four $\mu$TCA crates and Master Module.}
\label{figure_DAQMW_1Reader}
\end{figure}

DAQ-Middleware framework \cite{IEEE-Yasu} developed by KEK (The High Energy Accelerator Research Organization - Japan) allows easy development of DAQ software which runs on several PCs. 
A DAQ based on DAQ-Middleware was developed in previous ATCA system \cite{IEEE-Suzuki}. 
We reuse the DAQ-Middleware in our new $\mu$TCA system. 
The schematic view of DAQ-Middleware is shown in Figure \ref{figure_DAQMW_1Reader}. 
There are two components obtaining data: Fast Reader (reading data from $\mu$TCA system) and Slow Reader (reading slow control data such as High Voltage, temperature, etc.). 
For crate-parallel readout, we implement multiple threads inside Fast Reader. 
Fast Reader reads a full data set of one event. 
Event Builder is not introduced in the DAQ-Middleware although it is possible. 
With multiple threads, event building is done inside Fast Reader. 

\section{DAQ Performance}
DAQ performances are measured for:
\begin{itemize} 
\item{} multiple threads with multiple event buffers including single event readout time;
\item{} data taking efficiency of crate-parallel readout;
\item{} accepted rate of event-parallel readout.
\end{itemize}
For evaluating DAQ performance, the full setup of CANDLES (as shown in Figure \ref{figure_DAQ_diagram}) is used.
\par
In order to check the performance of multi-threads readout, a readout time measurement was conducted. 
In the case of single event buffer, dead-time is equivalent to readout time. 
The busy signal of Master Module indicates the dead-time of $\mu$TCA system. 
This busy signal is used in readout time measurement. 
This signal starts when global trigger sent and stops when all buffers in FPGA are free. 
In the measurement of readout time, we use a single buffer and 1 cps trigger rate from a function generator. 
A DSO7104B oscilloscope (Agilent Technologies \cite{Agilent-Tech}) with 4 Giga-samples per sec (GSa/sec) sampling speed is used to measure the width of the busy signal. 
The DAQ is configured with different number of threads and be checked readout time in every configuration. 
Since there are 4 crates in the DAQ, we configured 1, 2 and 4 threads in order to share data uniformly. 
The data rate is calculated with event data size mentioned in section \ref{AMC-FADCs}. 
Results of the measurement are shown in Table \ref{table_ReadTime}. 
The readout time is reduced with the increment of number of threads. 
With 4 threads, readout time is reduced 4 times: from 40 msec to 10 msec. 
This readout time is a half compared with previous ATCA DAQ system (20 msec \cite{IEEE-Suzuki}).

\begin{table}[!t]
\renewcommand{\arraystretch}{1.5}
\caption{Readout time of new $\mu$TCA system with different configuration of multi-threads for data readout}
\label{table_ReadTime}
\centering
\begin{tabular}{C{2.5cm} C{2.5cm} C{2.5cm}}
\hline
\hline
\bfseries Configuration & \bfseries Readout time/event & \bfseries Data Rate \\
\hline
1 thread & 40.4 $\pm$ 3.1 msec & 1.23 MB/sec\\
\hline
2 threads & 20.2 $\pm$ 1.2 msec & 2.45 MB/sec\\
\hline
4 threads & 10.1 $\pm$ 0.6 msec & 4.90 MB/sec\\
\hline
\hline
\end{tabular}
\end{table}

\begin{figure}
\centering
\includegraphics[width=80mm]{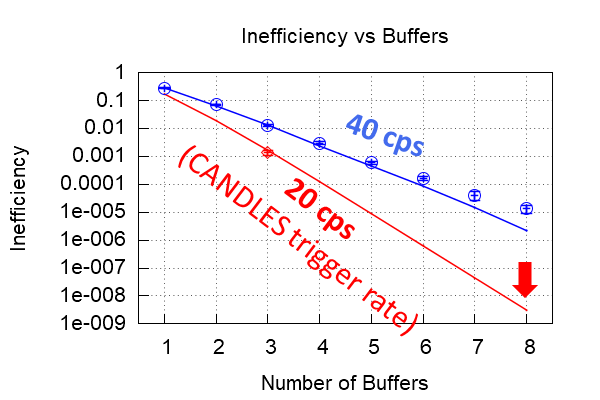}
\caption{Inefficiency as a function of multiple buffers. Inefficiency at 20 cps is calculated (red solid line). With 8 event buffers, inefficiency is very low (2.81$\times$10$^{-9}$), and it needs a long time to measure. To confirm, inefficiency at 40 cps is calculated (blue solid line) and measured (blue circle points). Measured and calculated data are consistent.}
\label{figure_Ineff_buffers}
\end{figure}

Event buffers in AMC-FADCs are functioned as derandomizer to reduce the dead-time. 
With readout time/event 10 msec, we can calculate inefficiency as a function of the number of event buffers. 
For 20 cps (CANDLES trigger rate), inefficiency with 8 event buffers is 2.81$\times$10$^{-9}$. 
To confirm this calculation with experimental data, it takes very long time of measurement (about 200 days of measurement). 
Thus, we compare calculation and measurement at 40 cps. 
We use a pseudo-random pulse generator of a 41-bit linear feedback shift register (LFSR) in the FPGA with trigger rate of 40 cps. 
Figure \ref{figure_Ineff_buffers} shows measured data at 40 cps and calculation data at 20 cps and 40 cps. 
Measured data and calculation data at 40 cps are consistent.

\begin{figure}
\centering
\includegraphics[width=80mm]{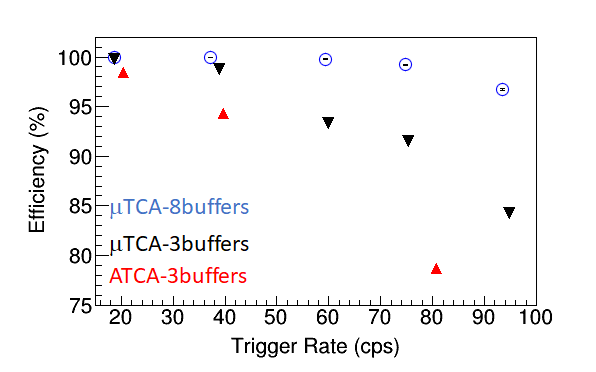}
\caption{DAQ data taking efficiency of $\mu$TCA with 8 buffers (red circle) and 3 buffers (black down-triangle). DAQ performance of $\mu$TCA system is compared with previous ATCA system (red up-triangle) \cite{IEEE-Suzuki}.}
\label{figure_DAQ_eff}
\end{figure}

High data taking efficiency is important to remove background in CANDLES. 
The data taking efficiency of DAQ system is measured as a function of trigger rate. 
A pseudo-random pulse generator with trigger rate from 20 cps to 100 cps is used. 
In this test, we use 4 threads and 8 buffers. 
In order to compare with previous ATCA system, we also did the measurement with 3 buffers configuration. 
Figure \ref{figure_DAQ_eff} shows the data taking efficiency. 
Data of $\mu$TCA system with 3 and 8 event buffers are plotted with black down-triangle and blue circle points, respectively. 
Data of ATCA system (with 3 event buffers and 3 PCs for data event-parallel readout) is plotted with red up-triangle points. 
Efficiency of $\mu$TCA with 3 buffers and 8 buffers are better than the ATCA system because the readout time of $\mu$TCA is a half of ATCA system. 
The $\mu$TCA DAQ with 4 threads and 8 buffers has the best performance. 
At trigger rate of CANDLES (about 20 cps \cite{IEEE-Suzuki}), we obtain no event lost after taking 4.2$\times$10$^{6}$ events (63 hours of data taking) with 4 threads and 8 buffers. 
The inefficiency is less than 10$^{-6}$, or the efficiency is very close to 100\%. 
At same trigger rate, ATCA system achieved 98\% to 99\% of efficiency. 
Thus, it proves that the $\mu$TCA has enough performance for CANDLES experiment. 

The event rate, which is mainly background rate, in the normal run is low. 
Therefore, throughput is not important, while high data taking efficiency is required. 
Beside normal run, calibration measurement with radioactivity source is carried out every 3 months for checking detector stability. 
In calibration run, we acquire data at high trigger rate (up to a few thousand events/sec). 
All event buffers are always occupied, so they cannot help to increase efficiency. 
Throughput in calibration run is, hence, important. 
Event-parallel was developed to speed up DAQ speed at high trigger rate. 
\par
\begin{figure}
\centering
\includegraphics[width=80mm]{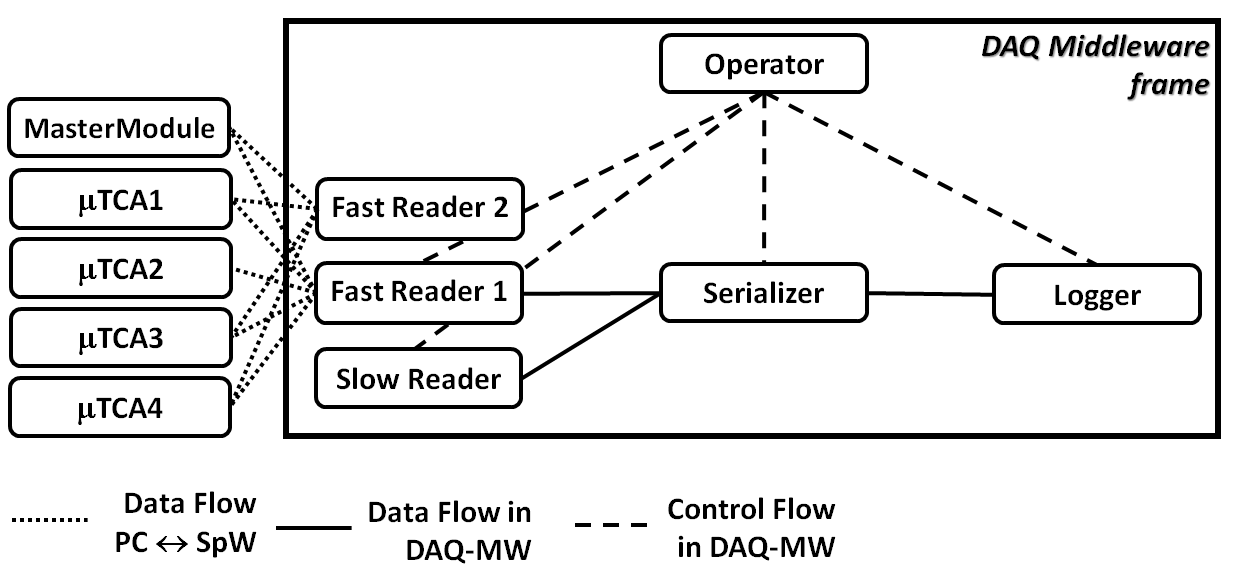}
\caption{Schematic diagram of DAQ-MW for event-parallel readout. 2 Fast Readers get data from event buffers of AMC-FADCs in $\mu$TCA system. Fast Reader 1 accesses even buffers, and Fast Reader 2 accesses odd buffers.}
\label{figure_DAQ_MW_2Reader}
\end{figure}
\begin{figure}
\centering
\includegraphics[width=80mm]{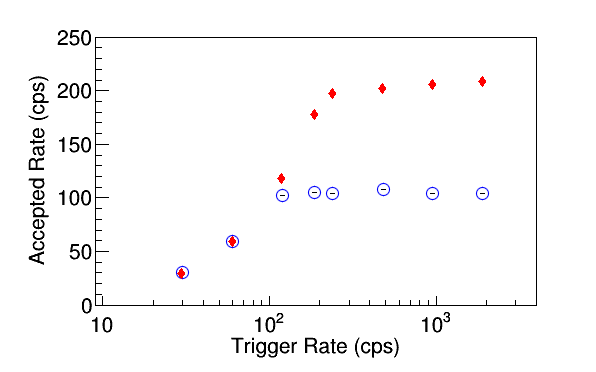}
\caption{Accepted Rate of one Fast Reader (blue circle) and two Fast Readers (red diamond). The readout rates of one Fast Reader and two Fast Readers are limited up to 100 cps and 200 cps, respectively.}
\label{figure_1Reader-vs-2Readers}
\end{figure}

For realizing event-parallel, there is nothing changed in hardware setup since we conduct 2 Fast Reader components in the same computer. 
Each Fast Reader component reads data with 4 threads accessing modules in 4 crates. 
Figure \ref{figure_DAQ_MW_2Reader} is the schematic diagram of DAQ-MW for event-parallel readout. 
These two Fast Readers share data readout in 8 buffers: one gets data from even buffers, while the other gets data from the odd buffers. 
Serializer component manages to arrange data of 8 buffers sequentially and adds data of Slow Reader. 
Serializer sends data to Logger for storage on the hard disk. Event Server and Monitor components are not prepared for this configuration. 
Since data taking time is short, we decided to do offline analysis after data taking. 
The performance was done with a pseudo-random function generator with trigger rate ranging from 30 cps to 2000 cps. 
Figure \ref{figure_1Reader-vs-2Readers} shows the accepted rate as a function of trigger rate with 1 Fast Reader (blue circle points) and 2 Fast Readers (red diamond points). 
Since readout time is 10 msec, accepted rate of 1 Fast Reader is limited at 100 cps at the trigger rate higher than 100 cps. 
On the other hand, two Fast Readers reduce readout time to 5 msec, maximum accepted rate is about 200 cps at trigger rate higher than 200 cps. 
We achieve DAQ speed two times faster with 2 event-parallel readouts. 

\section{SUMMARY}
A new $\mu$TCA DAQ system was introduced in CANDLES experiment. 
The new DAQ system includes AMC-FADCs with 8 event buffers and SpaceWire-GigabitEthernet interface. 
SpaceWire-GigabitEthernet helps us build a flexible network with off-the-shelf PC. 
However, it has overhead due to turn around time in the software. 
Since bandwidth utilization is low, there are rooms to improve by parallel readout. 
In this work, we handle the overhead issue by introducing parallel readout. 
There are 2 kinds of parallel readout applied in CANDLES DAQ system: crate-parallel readout and event-parallel readout. 
Crate-parallel with 4 multiple threads is introduced. 
With 4 threads, we can reduce the readout time 4 times (40 msec down to 10 msec). 
We conducted the measurement with 4 threads and 8 event buffers at CANDLES trigger rate in 63 hours and observed no event lost. 
Thus, the inefficiency of CANDLES is less than 10$^{-6}$ according to experimental data. 
In our calculation, the inefficiency is 2.81$\times$10$^{-9}$. 
After the passive shielding construction \cite{AIP-Nakajima} in CANDLES, the trigger rate is reduced to 10 cps. 
We expect lower inefficiency at 10 cps of trigger rate. 
At high trigger rate, dead-time increases and event buffers are always occupied. 
To increase data taking efficiency, we need the high throughput of event-parallel instead of event buffers. 
We set up 2 event-parallel readers getting data from 8 event buffers. 
The accepted rate is increased by 2 times: 100 cps to 200 cps. 
The data rate is increased from 5 MB/sec to 10 MB/sec. 
With the increment of DAQ speed, calibration measurement in CANDLES can be conducted in a shorter time and we can keep live time for 0$\nu\beta\beta$ study.

\end{document}